# Phase-Matching of High-Order Harmonics Driven by Mid-Infrared Light

Tenio Popmintchev,[1,*] Ming-Chang Chen,[1] Oren Cohen,[1] Michael E. Grisham,[2] Jorge J. Rocca,[2] Margaret M. Murnane,[1] and Henry C. Kapteyn[1]

*National Science Foundation ERC for Extreme Ultraviolet Science and Technology*

[1]*JILA, University of Colorado at Boulder and NIST, Boulder, CO 80309, USA*

[2]*Department of Electrical and Computer Engineering, Colorado State University, Fort Collins, CO 80523, USA*

*Corresponding author: popmintchev@jila.colorado.edu

We demonstrate that phase-matched frequency upconversion of ultrafast laser light can be extended to shorter wavelengths by using longer driving laser wavelengths. Experimentally, we show that the phase- matching cutoff for harmonic generation in argon increases from 45 to 100 eV when the driving laser wave- length is increased from 0.8 to 1.3 m. Phase matching is also obtained at higher pressures using a longer- wavelength driving laser, mitigating the unfavorable scaling of the single-atom response. Theoretical calculations suggest that phase-matched high harmonic frequency upconversion driven by mid-infrared pulses could be extended to extremely high photon energies.

High-order harmonic generation (HHG) is a unique source of femtosecond-to-attosecond duration soft x-ray beams that has made possible new studies of atoms, molecules, and materials, as well as enabling high-resolution coherent imaging using a tabletop source. To date, however, most applications of HHG radiation employ relatively low energy extreme ultraviolet wavelengths, because the efficiency of the HHG process decreases rapidly at high photon energies. This decrease is not fundamental to the nonperturbative HHG process, whose effective nonlinearity scales relatively slowly with photon energy, but rather results from the large phase mismatch between the harmonic field and the driving laser field. The difficulty in phase-matching high harmonic upconversion to very short wavelengths using a 0.8 μm driving laser is that the required high incident laser intensity strongly ionizes the medium, thus creating a large free electron dispersion that prevents phase matching [1]. True phase matching of the HHG process is thus limited to a parameter range where the levels of ionization of the medium is low, where neutral atom dispersion can balance the anomalous free-electron plasma dispersion [2, 3]. For a 0.8 μm driving laser, the "critical" ionization levels above which true phase matching is not possible are $\eta_{cr} \approx 5\%$ for argon and ≈0.5% for helium, corresponding to highest photon energies that can be phase matched of ~45 eV and ~130 eV respectively. Quasi-phase matching (QPM) techniques have been employed for generating short-wavelengths harmonics at ionization levels where true phase-matching is not possible [4-7]. However, by their nature, all QPM techniques only partially correct the phase mismatch and therefore are fundamentally less efficient than true phase matching. To date, QPM techniques have been implemented only over relatively short distances in the medium of ≈ 0.5mm.

Another important limit in HHG is the highest photon energy (cutoff energy) that can be generated by the laser, regardless of phase matching. This cutoff energy is given by $h\nu_{max} = I_p +$ 3.2 $U_p$, where $I_p$ is the ionization potential of the gas and $U_p \sim I_L \lambda_L^2$ is the quiver energy of the recolliding electron, $I_L$ is the laser intensity, and $\lambda_L$ is the wavelength of the driving laser. The favorable $\lambda_L^2$ scaling of the cutoff has motivated studies of HHG with mid-infrared driving pulses [8]. Significant extension of the cutoff energy was demonstrated in several experiments [8,9]. However, it was recently found theoretically that the single-atom yield scales as $\lambda_L^{-5.5\pm0.5}$ which greatly reduces the efficiency of HHG driven by longer wavelengths [13]. Thus, increasing the HHG yield by phase-matching the conversion process is critical to obtain a usable flux. Indeed, it was recently suggested theoretically that favorable "self phase matching" conditions might be realized with mid-IR pulses [11].

In this Letter, we show experimentally and theoretically that it is possible to extend true phase-matching of the high harmonic generation process to significantly higher photon energies (in theory to ≈ 1 keV) using long wavelength driving lasers, and that this phase matching process is consistent with our previously-established understanding of phase matching in HHG. Using a 1.3 µm driving laser and pressure-tuned phase matching in a hollow waveguide, we show that the region of phase matching using argon gas as the nonlinear medium is extended from ≈45 eV (0.8 µm driver) to ≈100eV (1.3 µm driver). Moreover, we show that phase matching is obtained at higher pressures using longer driving wavelengths, thereby helping to mitigate the above-mentioned unfavorable single-atom efficiency scaling. Most importantly, as a result of the ability to implement phase matching of the HHG process at high gas pressure, the high harmonic output from argon in the phase-matched region of 70-100 eV driven by 1.3 µm light is of comparable brightness to that of phase-matched He using 0.8 µm driving lasers. This is, to our knowledge, the first demonstration that mid-IR lasers can generate sufficient HHG flux for application experiments. Finally, we observe HHG emission up to 200 eV using 1.3 μm light - the highest photon energy generated to date using this driving laser wavelength.

Pressure-tuned phase-matched high harmonic generation involves creating a near plane-wave propagation geometry by proper coupling of high-intensity light inside a hollow waveguide [3]. In this geometry, the phase mismatch is a sum of the contributions from the pressure-dependent neutral and free electron plasma dispersion, as well as the pressure-independent waveguide dispersion:

$$\Delta k = q\left\{\left(\frac{u_{11}^2 \lambda_L}{4\pi a^2}\right) - P\left((1-\eta)\frac{2\pi}{\lambda_L}\Delta\delta - \eta N_{atm} r_e \lambda_L\right)\right\} \qquad (1)$$

In Eq. (1), $q$ is the harmonic order, $u_{11}$ is the lowest-order waveguide mode factor, $\lambda_L$ is the wavelength of the driving laser, $a$ is the inner radius of the hollow waveguide, $P$ is the pressure, $\eta$ is the ionization level, $r_e$ is the classical electron radius, $N_{atm}$ is the number density of atoms at 1 atm, and $\Delta\delta(\lambda_L)$ is the difference between the indices of refraction of the gas at the fundamental and harmonic wavelengths. Phase matching ($\Delta k=0$) and conversion efficiency can be optimized by simply tuning the gas pressure inside the waveguide. However, this is only possible as long as the ionization level is smaller than the critical ionization level, $\eta_{cr}$, given by $\eta_{cr}(\lambda_L) = [\lambda_L^2 N_{atm} r_e / (2\pi\Delta\delta(\lambda_L)) + 1]^{-1}$.

The critical level of ionization for phase matching results in a "phase matching cutoff energy" which is the largest photon energy that can be generated before the medium reaches the critical ionization level. This phase matching cutoff changes only slightly for shorter laser pulses and thus this cutoff represents a fundamental limit for conventional phase matched upconversion.

We first evaluate the dependence of the phase-matching cutoff energy on the wavelength of the driving laser by calculating the ionization level of the gas using the Amossov-Delon-Krainov tunneling model [12]. Here we assume a hyperbolic secant pulse of 8 cycles at FWHM (35 fs at 1.3 µm) with peak intensity selected such that the ionization level at the peak of the pulse corresponds to the wavelength dependent critical ionization. Figure 1 shows the result of this calculation, which indicates a significant increase in the maximum phase matched photon energy with driving laser wavelength. In-fact, for $\lambda_L$ approaching 2.8 µm, phase-matched conversion can be obtained even up to ~1 keV. Although the critical ionization level decreases as a function of $\lambda_L$ (since the longer wavelength driver experiences stronger free-electron dispersion), the pondermotive energy $U_p$ of the returning electron increases. Thus, higher energy photons are generated at lower laser intensities where the medium is much-less ionized. The result is that the phase matching cutoff (corresponding to harmonic generation below critical ionization is reached) moves to significantly higher energy.

Experimental data verifying the extension of the pressure-tuned phase-matching cutoff using long driving laser wavelengths is presented in Figure 2. High energy (up to 2.1 mJ), 35 fs pulses at 1.3 µm were generated through three-stage optical parametric amplification of a strongly chirped white light continuum. Conversion efficiencies of 28% were achieved for the signal at 1.3 µm (47% total efficiency for both signal and idler) when 7.5 mJ, 800 nm pump with 25 fs pulse duration was used. These pulses were then focused into a hollow waveguide of 250 µm inner diameter and 20 mm length filled with high pressure Ar. The gas was injected into the waveguide through two laser drilled holes, allowing a static gas pressure to be maintained within an interaction length of 10 mm, while two 5mm end sections enabled differential pumping to a vacuum of $10^{-3}$-$10^{-7}$ torr. The harmonic signal was detected using a flat field x-ray spectrometer (<2.5% estimated average transmission between 70 and 100 eV) and an x-ray CCD (Andor). Thin, 200 nm, Zr filters were used to block the driving laser mid-IR photons. Since the CCD is much less sensitive to 1.3 μm light compared with 0.8μm, fewer filters are required to eliminate the laser background. The K absorption edges of Al, Si and B were used for energy calibration of the harmonic spectra.

Figure 2(a) demonstrates that using a 1.3 μm driving laser, true phase-matching in Ar extends up to 100eV. In comparison, using a 0.8 μm driving laser, phase matching in Ar is limited to ≈ 45eV [1-3]. Based on the driving laser intensity of $2\times10^{14}$ W/cm$^2$ and pulse duration of 35 fs, ADK calculations show that the ionization level at which the single-atom cutoff reaches 100 eV is 0.6%, which is well below the critical ionization level of 1.5% for a 1.3 μm driving beam in Ar. Consistent with our finding, the phase matching cutoff energy at this laser wavelength is calculated to be 110 eV for a 35 fs pulse. Interestingly, this phase matching cutoff would increase by only 5% using a shorter, 20 fs, driving laser pulse. In Fig. 2, the harmonic spectrum as a function of pressure shows that higher harmonics are phase-matched at higher pressures - as is expected for pressure-tuned phase matching. This is because higher harmonic orders are generated at a higher level of ionization, and optimum phase matching occurs at a higher neutral atom pressures, to allow the neutral atom index to compensate for the increasing free-electron induced phase-mismatch, as seen from Eq. (1). An analytical model described in [3] was used to predict the ionization level at optimum pressure for efficient conversion, by fitting of the experimental data to the calculated growth of the harmonics in the presence of absorption. Ionization fractions of 1.0-1.3% and 1.2-1.3% were obtained for harmonics at 76 and 92 eV, respectively. In Fig. 2, the pressure tuned phase matching peaks are relatively broad, due to the presence of ionization in the medium, and well as radially and longitudinally varying laser intensity.

The hollow waveguide geometry is particularly useful in the case of a long-wavelength driving laser for several reasons. First, it makes it possible to obtain an optimal pressure-length product for phase matching. The observed phase matching peaks occur at higher pressures using 1.3 µm: 70-200 torr for a 1.3 µm driving laser compared with 10-50 torr for 0.8 µm [3]. These pressures would be very difficult to obtain in a free space jet or even a differentially pumped cell geometry. Second, guiding of the laser beam maintains high intensity over an extended interaction distance - even for a relatively long-wavelength driver (our experimentally measured waveguide transmission was > 70%). This is in contrast to a free focusing geometry where the confocal parameter of a Gaussian beam shrinks as $1/\lambda_L$ corresponding to $1/\lambda_L^2$ in intensity. Thus, although the $\lambda^{-5.5}$ scaling of the single atom efficiency with driving wavelength corresponds to an 11-18 times weaker emission for 1.3 µm compared with an 0.8 µm driving pulse, because of the ability to phase match the process, the photon flux can still be high. For a direct comparison, we used light at 0.8 µm to generate phase matched harmonics in He, but with approximately 2.6x higher peak laser intensity so that the same 130 eV cutoff photon energy was achieved. (Using 0.8 µm driving lasers, it is not possible to phase matching the HHG process in Ar at these photon energies). Figure 3 compares phase matched harmonic emission of Ar driven by 1.3 µm light at an intensity of $2 \times 10^{14}$ W/cm$^2$, with emission from He driven by 0.8 µm light at an intensity of $5.2 \times 10^{14}$ W/cm$^2$. The HHG flux from Ar driven by 1.3 µm beams is comparable or higher than that of the brightest source in the region – phase-matched He at 0.8 µm. At even higher IR laser intensities of $4.5 \times 10^{14}$ W/cm$^2$, HHG emission from argon driven by 1.3 µm light extends up to 200 eV (see Fig. 4), which is the maximum harmonic photon energy generated to date by this driving laser wavelength.

In summary, we demonstrate experimentally and theoretically for the first time that increasing the wavelength of the driving pulse in high harmonic generation significantly increases the cutoff energy for true phase matching. We demonstrate that pressure-tuned phase matching of the HHG process in the hollow waveguide geometry using longer wavelength drivers can accommodate the high pressures and long interaction lengths required for efficient upconversion. We demonstrate for the first time that phase matching from Ar at photon energies of 100 eV using a 1.3 µm driving pulse is as bright as emission from phase-matched He driven by 0.8 µm light. This result demonstrates a path for producing *bright* coherent x-rays for biological and materials imaging up to photon energies approaching 1 keV.

The authors gratefully acknowledge support from the U.S. Department of Energy and the NSF ERC for Extreme Ultraviolet Science and Technology under NSF Award No. 0310717.

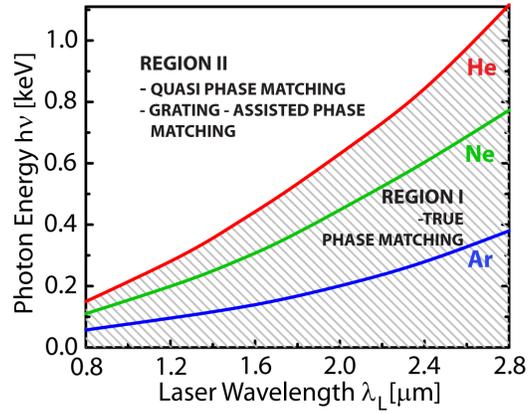

**Fig. 1.** Theoretically predicted "phase- matching cutoff energies" as a function of the laser wavelength $\lambda_L$ showing that true phase matching could be extended to 1 keV at ionization levels below $\eta_{cr}$ (region I). Quasi-phase matching or grating-assisted phase-matching techniques must be implemented at higher energies (region II).

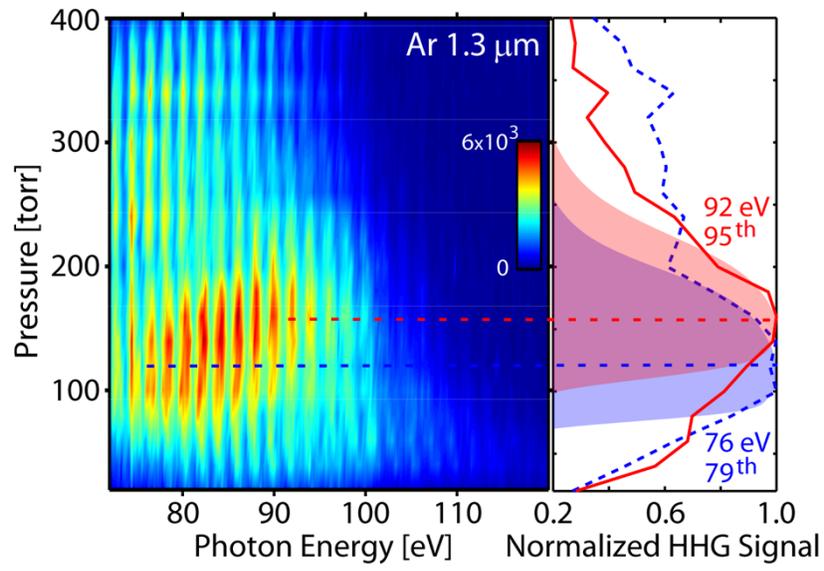

**Fig. 2.** (Left) Experimental HHG spectrum through 200 nm Zr filter as a function of pressure, demonstrating phase-matched emission from Ar up to 100 eV using a 1.3μm driving pulse. (Right) Lineouts showing the pressure dependence of the harmonics around 76eV (dashed line) and 92 eV (solid line) together with theoretical curves.

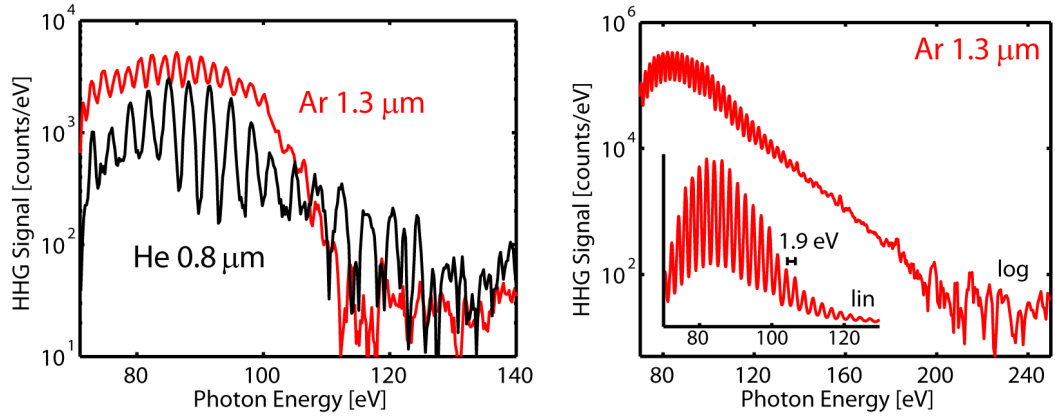

**Fig.3.** Comparison between HHG from phase-matched Ar (160 Torr, 2.0 x $10^{14}$ W/cm$^2$) driven by 1.3μm light and phase-matched He (500 Torr, 5.2 x $10^{14}$ W/cm$^2$) driven by 0.8μm light. (b) Harmonic spectrum from almost fully ionized Ar at lower pressure (60 Torr) driven by a 1.3 m pulse at an intensity of 4.5 x $10^{14}$ W/cm$^2$, showing emission up to 200 eV.